\documentclass[cits]{PoS}
\usepackage{wrapfig,rotating}
\usepackage{amssymb,amsmath,array}
\usepackage{fontenc,times,mathptmx,graphicx}

\usepackage{booktabs}

\title{D* (+jets) in Deep Inelastic Scattering and Photoproduction }

\ShortTitle{D* (+jets) in DIS and Photoproduction}

\author{\speaker{Andreas W. Jung}\thanks{Previously at Kirchhoff Institute of Physics, University of Heidelberg \& DESY, Hamburg.}~~(for the H1 collaboration)\\
        Fermi National Accelerator Laboratory (Fermilab)\\
        E-mail: \email{ajung@fnal.gov}}

\abstract{New results on charm quark production at HERA in an increased phase space in deep-inelastic
 scattering and photoproduction are discussed. Single \& double-differential cross section distributions
are compared to next-to-leading order QCD calculations as well as to MC@NLO in the photoproduction regime.
 The charm contribution to the proton structure, $F_2^{c\bar{c}}\left(x,Q^2\right)$, is determined with 
different experimental techniques and finally combined.}

\FullConference{35th International Conference of High Energy Physics - ICHEP2010,\\
		July 22-28, 2010\\
		Paris France}

\begin{document}

\section{Introduction}
HERA was the unique electron-proton ($ep$) machine colliding $27.5~\mathrm{GeV}$ electrons (positrons)
 with $920~\mathrm{GeV}$ protons providing a center-of-mass energy of $\sqrt{s} = 318~\mathrm{GeV}$
The charm quark production in $ep$ scattering is dominated by the boson-gluon-fusion (BGF) process 
$(\gamma p \rightarrow c\bar{c})$. This production process is directly sensitive to the gluon density 
in the proton and allows its universality to be tested. There are two different kinematic 
regions of charm quark production distinguished by the four-momentum transfer squared ($Q^2$) of the 
exchanged photon: The photoproduction regime with $Q^2 \lesssim 2~\mathrm{GeV^{2}}$ and the deep-inelastic 
scattering (DIS) regime with $Q^2 \gtrsim 5~\mathrm{GeV^{2}}$. Due to the presence of a hard scale 
($m_{c}, Q^2$ or $p_{_{T}}$) perturbative Quantum Chromodynamics (pQCD) can be applied. 
If one of the other scales is much bigger than the mass, charm quarks can be treated as massless ("massless scheme"), 
otherwise the mass needs to be taken into account ("massive scheme"). The latter assumes no charm quark content of the proton.

\section{$\mathbf{D^{*\pm}}$ Cross Section Measurements in Photoproduction \& DIS}
Events containing charm quarks are efficiently identified reconstructing $D^*$ mesons by the mass difference method. The new result \cite{photoPrel} in the photoproduction regime covers
 the kinematic region of $Q^2 < 2~\mathrm{GeV^2}$ and $0.1<y<0.8$ of the inelasticity of the scattering process. The $D^{*\pm}$ mesons have been tagged by the H1
Fast Track Trigger \cite{ftt_l3_proc}. They are selected if they have a $p_T (D^*) > \mbox{2.1}~\mathrm{GeV}$ and $|\eta (D^*)| < \mbox{1.5}$. The results make use of
 the full available data set corresponding to $L=93~\mathrm{pb^{-1}}$. Additional jets are reconstructed with the inclusive $k_T$ 
algorithm in the laboratory frame and are selected if they have a $p_T (Jet) > \mbox{3.5}~\mathrm{GeV}$. With the additional constraints 
given by the jets the reconstruction of the longitudinal momentum fraction of the photon $x_{\gamma}$ taking part in the hard interaction is possible. 
\begin{figure}[ht]
    \begin{minipage}[c]{0.47\columnwidth}
    \centerline{\includegraphics[width=1.\columnwidth]{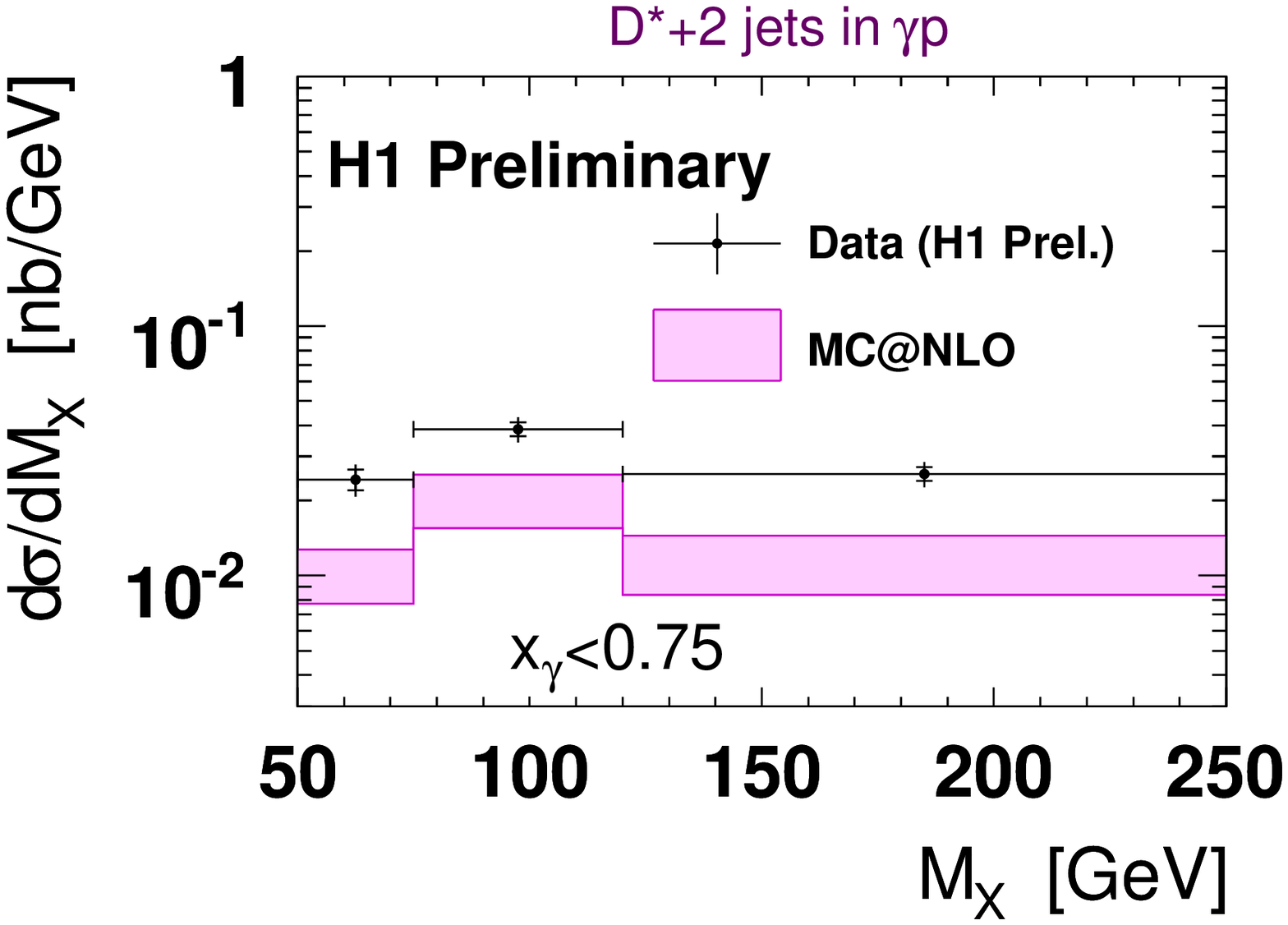}}
     \end{minipage}
     \hspace{.05\linewidth}
     \begin{minipage}[c]{0.47\columnwidth}
    \centerline{\includegraphics[width=1.\columnwidth]{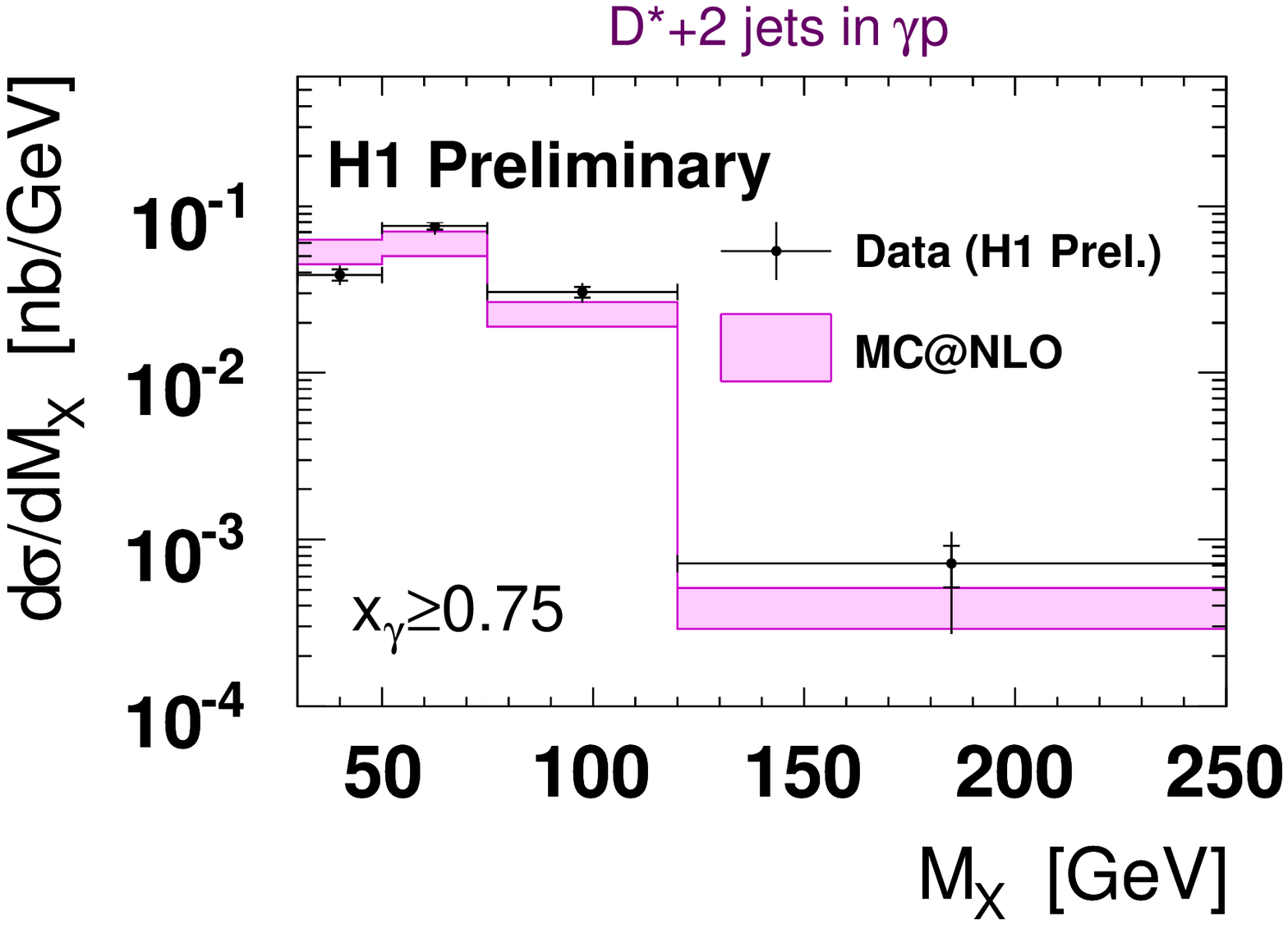}}
     \end{minipage}
  \caption{\label{fig:photoProd} $D*$ + 2 jet cross section as a function of the invariant mass of the proton and photon side $M_X$ for low 
$x_{\gamma} < 0.75$ (a) and high $x_{\gamma} > 0.75$ (b). The data are compared to the prediction by MC@NLO \cite{mcatnlo} using the CTEQ66 proton PDF \cite{cteq66}.}    
 \end{figure}
Thus the phase space can be divided into a region enriched with resolved photons ($x_{\gamma} < 0.75$), where a parton of the photon interacts with the proton
 and a region dominated by direct processes ($x_{\gamma} > 0.75$). It is found that the invariant mass $M_X$ of the proton and photon remnants for the direct 
region is described by MC@NLO (Fig. \ref{fig:photoProd}(b)), while the resolved part is too low in normalization (Fig. \ref{fig:photoProd}(a)). Especially low 
$x_{\gamma}$ is sensitive to the photon parton density function (PDF).\\
$D^*$ production in DIS has been measured for $0.02<y<0.7$ in two regimes of photon virtuality: $5<Q^2<100~\mathrm{GeV^2}$ and $100<Q^2<1000~\mathrm{GeV^2}$. 
Both analyses use the full HERAII data set corresponding to an integrated luminosity of about $350~\mathrm{pb^-1}$. In the lower $Q^2$ region the visible 
phase space  
\begin{figure}[ht]
    \begin{minipage}[c]{0.47\columnwidth}
    \centerline{\includegraphics[width=0.95\columnwidth]{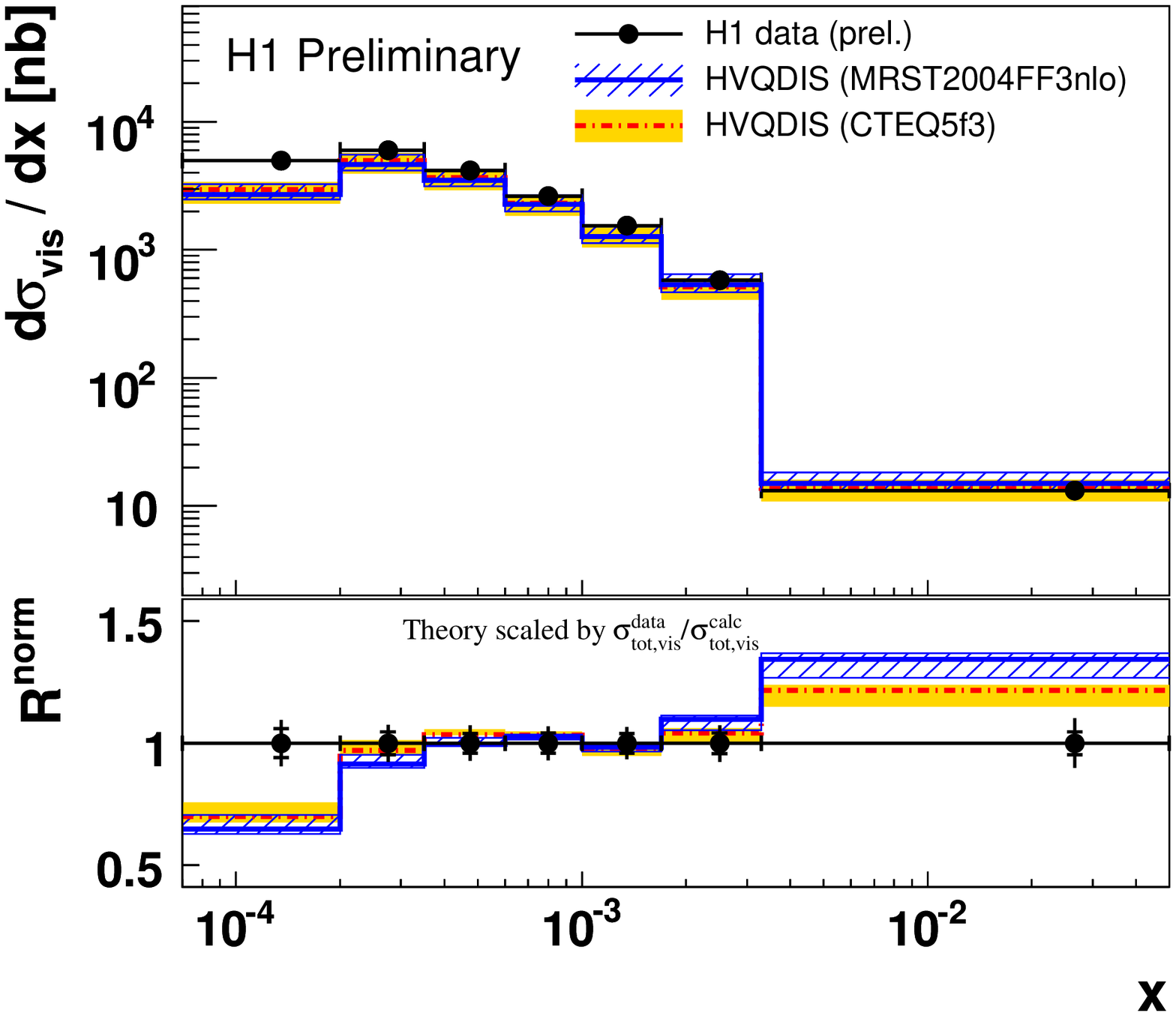}}
     \end{minipage}
     \hspace{.05\linewidth}
     \begin{minipage}[c]{0.47\columnwidth}
    \centerline{\includegraphics[width=0.95\columnwidth]{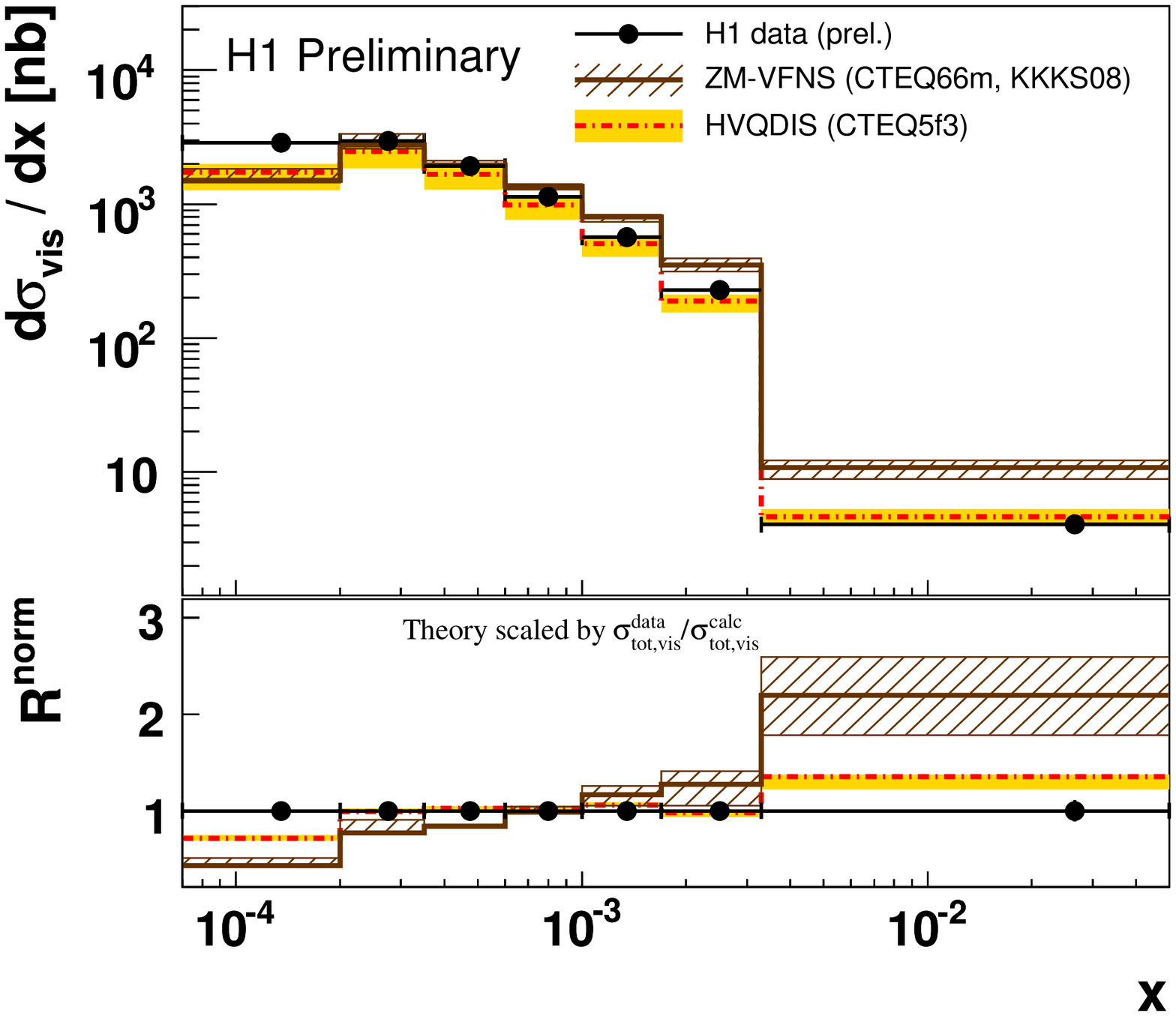}}
     \end{minipage}
  \caption{\label{fig:DIS_xsec} $D*$ cross section as a function of $x$ (a) compared to the NLO QCD calculation (HVQDIS) using two different proton
PDFs: MRST2004FF3nlo \cite{mrst04ff} or CTEQ5f3 \cite{cteq5f3}. (b) shows the $D*$ data for $p_T^* > 2~\mathrm{GeV}$ compared to the massless 
(ZM-VFNS) and massive NLO QCD calculation.}    
 \end{figure}
of the $D^*$ meson is restricted to $p_T (D^*) > \mbox{1.25}~\mathrm{GeV}$ and $|\eta (D^*)| < \mbox{1.8}$ \cite{medQ2Prel}. This measurement yields 
currently the largest phase space coverage at HERA for an inclusive $D^*$
 \begin{wrapfigure}{r}{0.5\columnwidth}
   \centerline{\includegraphics[width=0.495\columnwidth]{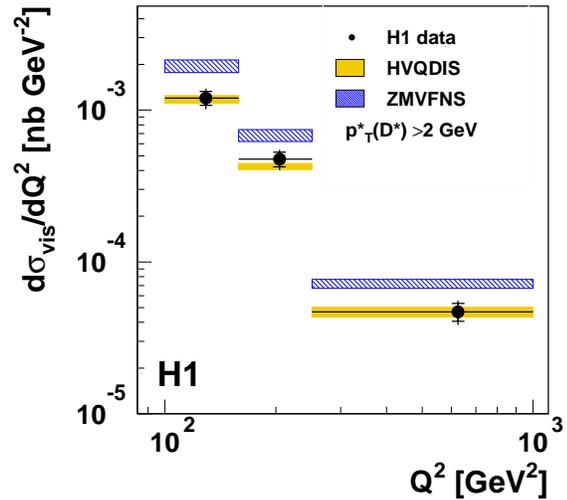}}
  \caption{\label{fig:highQ2_ZMVFNS} $D*$ cross section at high $Q^2$ compared to the massive (HVQDIS) and massless (ZM-VFNS) NLO QCD calculation.}    
 \end{wrapfigure}%
 cross section measurement. The Data are reasonably well described using the NLO calculation HVQDIS \cite{hvqdis} in the massive scheme 
(FFNS). However the slope in Bj\o rken $x$ is 
not well described (Fig. \ref{fig:DIS_xsec}a)).  An additional transverse momentum cut in the photon-proton rest frame of $p_{T}^* (D^*) > 2~\mathrm{GeV}$ is applied. 
In order to allow comparisons with the massless NLO calculation (ZM-VFNS) \cite{zmvfns} using the CTEQ66 proton parton density function (PDF) together
 with the fragmentation function KKKS08 \cite{zmvfns}. The massless calculation fails completely to describe the $x$ distribution (Fig. \ref{fig:DIS_xsec}(b)), 
whereas the data are reasonably well described by the massive NLO QCD calculation provided by HVQDIS.\\
The $D^*$ cross section result at high $Q^2$ \cite{highQ2Publ} is restricted to a visible phase space of $p_T (D^*) > \mbox{1.5}~\mathrm{GeV}$ and
 $|\eta (D^*)| < \mbox{1.5}$. It is found that the NLO calculation HVQDIS describes the data reasonably well (Fig. \ref{fig:highQ2_ZMVFNS}). An additional 
cut $p_T^* (D^*) > 2~\mathrm{GeV}$ has been applied to compare with the massless NLO calculation ZM-VFNS. Like in the medium $Q^2$ regime it fails to 
describe the data (Fig. \ref{fig:highQ2_ZMVFNS}). 

\begin{figure}[ht]
    \centerline{\includegraphics[width=0.95\columnwidth]{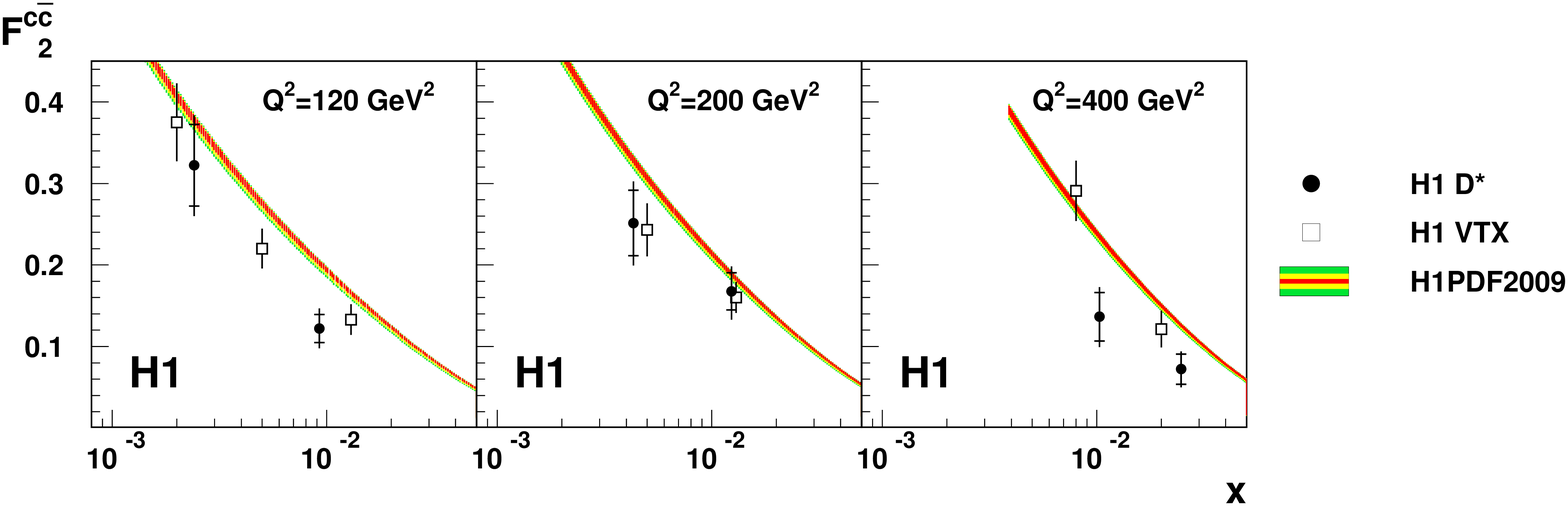}}
  \caption{\label{fig:highQ2_f2c} Charm contribution to the proton structure from $D^*$ cross sections together with the H1 results using lifetime 
information. Data are compared to the H1 PDF fit H1PDF2009.}    
 \end{figure}

\section{Measurement \& Combination of $\mathbf{F_2^{c\bar{c}}(x,Q^2)}$}
Different experimental techniques can be applied to measure the charm contribution, $F^{c\bar{c}}_2(x,Q^2)$, 
\begin{wrapfigure}{r}{0.555\columnwidth}
  \centerline{\includegraphics[width=0.55\columnwidth]{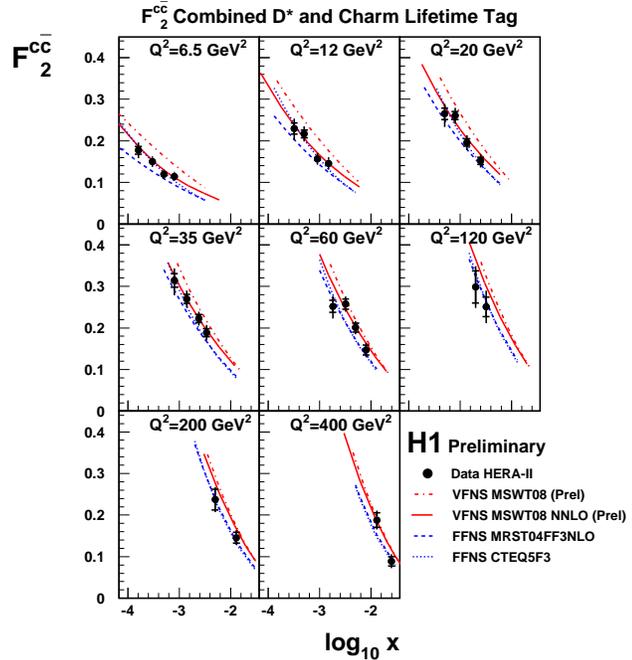}}
  \caption{\label{fig:allQ2_f2cComb} The H1 combined $F_2^{c\bar{c}}$ is compared to different NLO QCD calculations.}    
\end{wrapfigure}%
to the proton structure. Either the inclusive $D^*$ cross section measurements, semi-leptonic decays or the inclusive lifetime technique where the 
precise track information of the H1 
Silicon Vertex Detector is deployed. However, both techniques rely on the extrapolation of the measured cross sections to the full phase space. 
Published H1 results \cite{highQ2Publ} of $F^{c\bar{c}}_2(x,Q^2)$ at high $Q^2$ (Fig. \ref{fig:highQ2_f2c}) are reasonable described by the H1 PDF fit 
H1PDF2009 \cite{h1pdf2009}. The data are also in agreement with the H1 inclusive lifetime results \cite{lifetimePubl}. An improvement in the precision 
of the data is obtained by the combination of different $F^{c\bar{c}}_2(x,Q^2)$ measurements. The results at high $Q^2$ and a H1 
Preliminary result on $F^{c\bar{c}}_2(x,Q^2)$ at medium $Q^2$ \cite{jungF2c} are combined with a published measurement using variables reconstructed with the 
H1 Silicon Vertex Detector \cite{lifetimePubl}. This measurement uses a dataset corresponding to $L=189~\mathrm{pb^{-1}}$. The typical reduction of the 
systematic error is 25\%. The combined data \cite{f2cCombPrel} are compared to various NLO QCD calculations using different proton 
PDFs. The data are reasonably described by the different NLO QCD calculations in the FFNS scheme or the VFNS scheme \cite{vfns} which implements 
a transition from the massive to the massless charm treatment. For the FFNS scheme the 
proton PDFs MRST2004FF3nlo and CTEQ5f3 are used, whereas for the VFNS scheme the MSTW08 at NLO and NNLO are used. Especially at low $Q^2$ the data 
are precise enough to distinguish between models.

\section{Conclusions}
New measurements using the full H1 HERAII data sample have been analyzed for charm quark production in $ep$ scattering. Data in photoproduction are 
reasonable described by MC@NLO. The measurement in the DIS regime at medium $Q^2$ has been performed in the largest phase space at HERA for these 
kind of measurements. The data are reasonable described by the massive NLO QCD calculation. The massless 
NLO QCD calculation fails completely to describe the $x$ slope. The data at high $Q^2$ show a reasonable description by the NLO QCD calculations 
except for the massless NLO calculation which again fails to describe the data. The charm contribution to the proton structure at high $Q^2$ has 
been presented and is reasonably well described by the H1 PDF fit H1PDF2009. Finally the combination of the $F^{c\bar{c}}_2$ measurements from the
 $D^*$ results and the lifetime result was presented. The combined $F^{c\bar{c}}_2$ gives a more precise result and is reasonably described by different 
NLO QCD predictions. The data shows a sensitivity to the gluon density and at low $Q^2$ the data are precise enough to differentiate between models.

\end{document}